\documentclass[]{aa}

\usepackage[dvips]{graphicx}
\usepackage{amsmath}
\usepackage{amssymb}

\newcommand{\doh}{\mbox{${\rm 12+\log(O/H)}$}}

\begin{document}

\title{Simulating the mass-metallicity relation from 
$z\sim 1$\thanks{Research undertaken as part of the
Commonwealth Cosmology Initiative (CCI:www.thecci.org).}}

\author{M. Mouhcine\inst{1}
     \and
        B.~K. Gibson\inst{2,3}
     \and
        A. Renda\inst{4}
     \and
        D. Kawata\inst{4,5} }

\institute{
Astrophysics Research Institute, Liverpool John Moores University, 
Twelve Quays House, Egerton Wharf, Birkenhead, CH41~1LD, UK 
\and
Centre for Astrophysics, University of Central Lancashire, 
Preston, PR1~2HE, UK
\and
School of Physics, University of Sydney, NSW, 2006, Australia
\and
Centre for Astrophysics \& Supercomputing, Swinburne University, 
Hawthorn, Victoria, 3122, Australia
\and
The Observatories of the Carnegie Institution of Washington, 
813 Santa Barbara Street, Pasadena, CA, 91101 USA
}

\authorrunning{M. Mouhcine et~al.}

\abstract
% Context
{
The chemical properties of galaxies and their evolution as 
a function of cosmic epoch are powerful constraints of their 
evolutionary histories.
}
% Aims
{
This work provides a grid of numerical models of galaxy 
evolution over an extended cosmic epoch. The aims are to 
assess how well current models reproduce observed properties 
of galaxies, in particular the stellar mass versus gas phase 
metallicity relation, and to quantify the effect of the 
merging histories of galaxies on their final properties.
}
% Method
{
We use 112 N-body/hydrodynamical simulations in the 
standard Cold Dark Matter universe, to follow the formation 
of galaxy-sized halos and investigate the chemical enrichment 
of both the stellar component and the interstellar medium of 
galaxies, with stellar masses larger than $\sim10^9$~M$_{\odot}$. 
}
% Results
{
The resulting chemical properties of the simulated galaxies 
are broadly consistent with the observations. 
The predicted relationship between the mean metallicity and 
the galaxy stellar mass for both the stellar and the gaseous 
components at $z=0$ are in agreement with the relationships 
observed locally. The predicted scatter about these 
relationships, which is traced to the differing merging 
histories amongst the simulated galaxies with similar final 
masses, is similar to that observed. Under the hierarchical 
formation scenario, we find that the more massive galaxies 
are typically more evolved than their low mass counterparts 
over the second half of the age of the Universe. 
The predicted correlations between the total mass and the 
stellar mass of galaxies in our simulated sample from the 
present epoch up to $z\sim 1$ agree with observed ones. 
We find that the integrated stellar populations in the 
simulations are dominated by stars as old as $4-10$~Gyr. 
In contrast with massive galaxies, for which the 
luminosity-weighted ages of the integrated stellar populations 
in the simulated sample agree with those derived from the 
modeling of observed spectral energy distributions, simulated 
galaxies with stellar masses $\sim10^{9}$~M$_{\odot}$ at 
$z=0$ tend to be older than the local galaxies with similar 
stellar masses.
}
% Conclusions
{
The stellar mass versus metallicity relation and its 
associated scatter are reproduced by the simulations as  
consequences of the increasing efficiency of the conversion 
of gas into stars with stellar mass, and the differing 
merging histories amongst the galaxies with similar masses. 
The old ages of simulated low mass galaxies at $z=0$, and 
the weak level of chemical evolution for massive galaxies 
suggest however that our modelling of the supernova feedback 
may be incomplete, or that other feedback processes have 
been neglected.
}

\keywords{Methods: numerical -- Galaxies: abundances --
Galaxies: evolution -- Galaxies: formation }

\maketitle

\section{Introduction}

The chemical enrichment histories of galaxies provide insight 
into various processes involved in galaxy formation and 
evolution. The chemical composition of stars and gas within 
a galaxy depends on a number of physical processes, such as 
the star formation history, gas outflows and inflows, stellar 
initial mass function, etc. Although it is a complicated 
task to disentangle the effects of these processes, 
the galactic chemical abundances at various epochs place tight 
constraints on the likely evolutionary histories of galaxies.

The correlation between galaxy metallicity and luminosity 
in the local universe is one of the most significant 
observational results in galaxy chemical evolution studies. 
Lequeux et al. (1979) first revealed that the oxygen 
abundance increases with the total mass of irregular galaxies. 
The luminosity-metallicity relation for irregulars was later 
confirmed by Skillman et al. (1989) and Richer \& McCall 
(1995), amongst others. Subsequent studies have extended the 
relation to spiral galaxies (e.g., Garnett \& Shields 1987; 
Zaritsky et al. 1994; Garnett et al. 1997; Pilyugin \& Ferrini 
2000), and to elliptical galaxies (Brodie \& Huchra 1991). 
The luminosity correlates with metallicity over $\sim10$ 
magnitudes in luminosity and a factor of $\sim100$ in 
metallicity, with indications suggesting that the relationship 
may be independent of environment (Vilchez 1995; Mouhcine et al. 
2007) and morphology (Mateo 1998). More recently, large 
samples of star-forming galaxies drawn from galaxy redshift 
surveys, e.g. 2dF Galaxy Redshift Survey and the Sloan Digital 
Sky Survey (SDSS hereafter), have been used to confirm the 
existence of the luminosity-metallicity relation over a broad 
range of luminosity and metallicity (Lamareille et al. 2004; 
Tremonti et al. 2004). Lee et al. (2006) have extended the 
mass-metallicity relation to dwarf irregular galaxies, and 
found that the dispersion is similar over five orders of 
magnitude in stellar mass, and that the relation between 
the integrated stellar mass and the oxygen abundance of the 
interstellar medium is similarly tight from high stellar 
mass to low stellar mass galaxies.
{\bf
The gas phase oxygen abundance vs. stellar mass has been
understood either as a depletion sequence or a sequence in 
astration (see e.g. Tremonti et al. 2004). K\"oppen, Weidner, 
\& Kroupa (2007) have presented an alternative explanation 
of the mass-metallicity relation arguing it could be the 
consequence of the integrated galactic initial mass function 
depending on the star formation rate, leading to higher 
oxygen yield in systems where the star formation rate is high. 
More massive galaxies are expected to have higher yields and 
thus have a tendency to have higher metallicities.
Dalcanton (2007) presented a series of closed-box chemical 
evolution models including infall and outflow. She has shown 
that neither simple infall nor outflow can reproduce the 
observed low effective yields in low-mass galaxies (see 
Lee et al. 2006 for a similar conclusion), but metal-enriched 
outflows can do. That is, effectively the freshly synthesized 
elements need to be removed from the matter cycle, which is, 
as noted by K\"oppen et al. (2007) in principle, equivalent 
to reducing the number of massive stars.}
%
%The stellar populations of local low stellar mass galaxies 
%are found to be on average metal-poor with young 
%luminosity-weighted ages, whereas the massive galaxies are 
%found to be old and metal-rich, with a rapid transition 
%between these regimes (Gallazzi et al. 2005). 
%The mean metallicities and ages of the integrated stellar 
%populations show a large intrinsic scatter at a given 
%stellar mass indicating that the ages and metallicities 
%are not uniquely determined by the stellar mass.

Many recent studies in galaxy evolution trace changes in 
scaling relations of galaxies at earlier epochs. In this 
context, the galactic chemical abundances at different 
cosmic epochs can assist in constraining the likely 
scenarios of galaxy evolution. Different groups have used 
the classical nebular diagnostic techniques developed to 
study the properties of {H{\sc ii}} regions and emission 
line galaxies in the local universe to probe the properties 
of the interstellar gas in intermediate ($0<z<1$: 
Hammer et al. 2001; Lilly et al. 2003; Liang et al. 2006; 
Maier et al. 2004; Kobulnicky \& Kewley 2004; Maier et al. 
2005; Mouhcine et al. 2006ab; Lamareille et al. 2006) and 
high-redshift galaxies ($1.5<z<4$: Pettini et al. 1998; 
Kobulnicky \& Koo 2000; Mehlert et al. 2002; 
Lemoine-Busserolle et al. 2003; Erb et al. 2006; Maier et al. 
2006). The oxygen abundances for the interstellar medium for 
luminous star-forming galaxies at intermediate redshift are 
found to cover the same range as their local counterparts, 
and most of them fall on the local luminosity-metallicity 
relation. However, a minority appear to have significantly 
lower oxygen abundances than the local galaxies with similar 
luminosities (Kobulnicky et al.2003; Lilly et al. 2003; 
Liang et al. 2004; Mouhcine et al. 2006a). These luminous, 
massive, intermediate redshift, star-forming galaxies with 
low oxygen abundances have bluer colours than the higher 
metallicity ones, and exhibit physical conditions, i.e., 
emission line equivalent width and ionization state, very 
similar to those of the local faint and metal-poor 
star-forming galaxies (Mouhcine et al. 2006a; Maier et al. 
2005). The intermediate redshift massive and large galaxies 
with low gas-phase oxygen abundances are most likely immature
galaxies that will increase their metallicities and their
stellar masses to the present epoch. The diversity of the 
properties of the massive and large galaxies at intermediate 
redshifts supports the scenario whereby galaxies are still 
assembling their baryonic content between $z\sim 1$ and $z=0$ 
(Hammer et al. 2005). For a sample of mostly disk, intermediate 
redshift field and cluster galaxies, Mouhcine et al. (2006a) 
have shown that oxygen abundances do not correlate with either 
the maximum rotation velocity, i.e. a proxy of the total mass, 
or with the emission scale length size of the galaxy (see also 
Lilly et al. 2003). 
{\bf Savaglio et al. (2005) have shown that the observed 
redshift evolution of the mass-metallicity relation could be 
well reproduced by a simple closed-box model where the star 
formation timescale is proportional to the galaxy total 
baryonic mass. }

As these observational data for more statistically significant 
samples of galaxies accumulate, numerical simulations become 
important to understand galaxy formation in the context of 
the currently favored hierarchical clustering scenario.
To be compared with the observational data -- which provides
not only dynamical information, such as stellar and gas mass 
and rotation velocity, but also properties of stellar content,
such as metallicity and age -- such simulations are required 
to follow both the dynamical and chemical evolution of galaxies
(e.g.: Mosconi et al. 2001; Lia et al. 2002; Kawata \& Gibson 
2003; Kobayashi et al. 2006). Using chemodynamical numerical 
simulations, we have built an ensemble of simulated late-type 
galaxies, spanning a factor of 50 in total mass, and sampling 
a range of assembly histories at a given total mass. 
We compare our large sample of simulated galaxies with recent 
observational results, and examine how the current simulations
can explain the observations. We pay particular attention to 
the observed correlations between the galaxy mass and
metallicity of both the gaseous and stellar components, key 
observables in disentangling the formation history of galaxies.
In Section~\ref{simul} we describe our numerical simulations.
Section~\ref{resl} provides comparisons between simulated 
galaxies and observations. We summarize our conclusions in 
Section~\ref{concl}.

\section{Simulations}
\label{simul}

The ensemble of simulated galaxies analysed here are patterned 
after the adiabatic feedback model of Brook et al. (2004), 
using the chemo-hydrodynamical evolution code {\tt GCD+} 
(Kawata \& Gibson 2003). The simulation models based on the
semi-cosmological version of the code have been extensively 
used in our previous studies of galaxy properties. Brook et al. 
(2004) demonstrated that newly introduced adiabatic feedback 
model helps to create more realistic late-type disk galaxies. 
Brook et al. (2004; 2005) could reproduce the properties of 
both the thin and thick disk observed in the Milky Way and 
galaxies beyond the Local Group. Renda et al. (2005) used a 
part of the sample analysed here, to show that these simulated 
galaxies can reproduce the observed halo metallicities for 
galaxies with different masses. Details of both {\tt GCD+} 
and the feedback model can be found elsewhere 
(Kawata \& Gibson 2003; Brook et al. 2004; Renda et al. 2005), 
and we therefore only summarize briefly this information here.

The code is based on {\tt TreeSPH} (Hernquist \& Katz 1989; 
Katz, Weinberg \& Hernquist 1996), which combines the tree 
algorithm (Barnes \& Hut 1986) for the computation of the 
gravitational forces with the smoothed particle hydrodynamics 
(SPH) (Lucy 1977; Gingold \& Monaghan 1977) approach to 
numerical hydrodynamics. The dynamics of the dark matter 
and stars is calculated by the $N$-body scheme, and the 
gas component is modeled using SPH. It is fully Lagrangian, 
three-dimensional, and highly adaptive in space and time 
owing to individual smoothing lengths and individual time 
steps. Moreover, it includes self-consistently almost all 
the important physical processes in galaxy formation, such 
as self-gravity, hydrodynamics, radiative cooling, star 
formation, supernova feedback and metal enrichment.

Radiative cooling is computed using a metallicity-dependent
cooling function (derived with MAPPINGSIII: Sutherland 
\& Dopita 1993). The cooling rate for solar metallicity gas
is larger than that for gas of primordial composition by 
more than an order of magnitude. Thus, the cooling by metals 
should be accounted for in numerical simulations of galaxy 
formation. We use the following three criteria for star 
formation: (i) the gas density is greater than a critical 
density, 
$\rho_{\rm crit} = 2 \times 10^{-25}\ {\rm g\ cm^{-3}}$,
i.e.\ $n_{\rm H} \sim 0.1 {\rm cm^{-3}}$;
(ii) the gas velocity field is convergent,
${\bf \nabla} \cdot \mbox{\boldmath $v$}_i < 0$; and 
(iii) the Jeans unstable condition, $h/c_s>t_{\rm g}$, 
is satisfied, here $h$, $c_s$, and 
$t_{\rm g} = \sqrt{3 \pi/16 G \rho_{\rm g}}$ are the SPH 
smoothing length, the sound speed, and the dynamical time 
of the gas respectively. We used a fixed star formation 
efficiency independently of the halo mass. 
The effect of the ultraviolet background radiation field was 
not included.

We distribute the feedback energy from supernovae explosions 
in purely thermal form, although a fraction of it could, in 
principle, be distributed in kinetic form. 
Following Brook et al. (2004), the energetic feedback is 
implemented such as the gas within the SPH smoothing kernel 
of Type~II supernovae is prevented from cooling for the 
lifetime of the lowest mass star that ends as a Type~II 
supernovae, i.e. the lifetime of an 8~M$_{\odot}$ star (see 
also Thacker \& Couchman 2000). 

The metal enrichment was derived using supernovae Type~II 
and Ia, and the mass loss from intermediate mass stars, 
relaxing the instantaneous recycling approximation. The code 
calculates the event rates of SNe II and SNe Ia, and the yields 
of SNe II, SNe Ia and intermediate mass stars for each star 
particle at every time step, considering the Salpeter (1955) 
initial mass function (mass range of 
0.1-60 ${\rm M_{\rm \sun}}$) and metallicity dependent 
stellar lifetimes. We assume that each massive star 
($\geq8\ {\rm M_{\rm \sun}}$) explodes as a Type II supernova.
The SNe Ia rates are calculated using the model proposed by 
Kobayashi, Tsujimoto \& Nomoto (2000). The yields of SNe II, 
SNe Ia and intermediate mass stars are taken from Woosley 
\& Weaver (1995), Iwamoto et al. (1999), and van den Hoek 
\& Groenewegen (1997). The simulation follows the evolution 
of the abundances of several chemical elements ($^1$H, $^4$He,
$^{12}$C, $^{14}$N, $^{16}$O, $^{20}$Ne, $^{24}$Mg, $^{28}$Si, 
$^{56}$Fe, and Z, where Z is the total metallicity).

For each model, we start with an isolated sphere of dark 
matter and gas, on to which small-scale density fluctuations 
based on a CDM power spectrum are superimposed using
\texttt{COSMICS} (Bertschinger 1998). These fluctuations are 
the seeds for local collapse and subsequent star formation. 
The ``top-hat'' overdensity has an amplitude $\delta_{\rm i}$ 
at initial redshift $z_{\rm i}$, which is approximately 
related to the collapse redshift $z_{\rm c}$ by 
$z_{\rm c}=0.36\delta_{\rm i}(1+z_{i})-1$ 
(e.g. Padmanabhan 1993). Solid-body rotation corresponding 
to a spin parameter $\lambda$ is imparted to the initial 
sphere to mimic the effects of longer wavelength 
fluctuations. As we focus here on the effect of the
differences in galaxy assembly histories on galaxy properties,
we fixed the collapse redshift and spin parameter. 
To guarantee that the end-product of the simulations would be 
disk-like system, we have chosen a high spin parameter (see 
also Brook et al. 2004; 2005). The relevant parameters include 
$\Omega_{0}=1$, baryon fraction $\Omega_{\rm b}=0.1$, 
$\sigma_{8}=0.5$, $H_{0}=50$~km~s$^{-1}$~Mpc$^{-1}$, 
the collapse redshift $z_{\rm c}=2$, and the spin parameter 
$\lambda=0.060$. 

A series of 112 simulations were generated using the same 
collapse redshift $z_{\rm c}$, and spin parameter $\lambda$, 
for four different halo total masses, 
${\rm M_{tot}~=~10^{11}~M_{\odot}}$,
${\rm 5\times10^{11}~M_{\odot}}$,
${\rm 10^{12}~M_{\odot}}$, and
${\rm 5\times10^{12}~M_{\odot}}$. 
We used 14147 dark matter and 14147 gas/star particles, for 
the ensemble of simulations analysed here. For each total 
mass, we run models with different patterns of initial 
small-scale density fluctuations which lead to different 
hierarchical assembly histories. This was controlled by 
setting different random seeds for the Gaussian perturbation 
generator in {\tt COSMICS}. 

In our main sample, both the collapse redshift and the spin 
parameter have been fixed in order to sample only the pattern 
of initial density perturbations, and thus the formation 
history. To investigate the properties of galaxies which 
collapsed at a lower redshift, we also run a subset of 
9 simulations with a total mass of ${\rm 10^{11}~M_{\odot}}$,
collapse redshift $z_{\rm c}=1.5$, and spin parameter 
$\lambda=0.054$. Note that the definition of collapse 
redshift is for a spherical system with no rotation. 
If there is rotation, i.e., additional kinetic energy, 
the real collapse redshift would be lower and it is more 
complicated to be estimated. Therefore, we have chosen a 
lower spin parameter for this lower collapse model, to 
guarantee that the system collapses before z=0.

Admittedly, the semi-cosmological models are not fully 
self-consistent galaxy formation models, compared to the
full-cosmological simulations (e.g.: Steinmetz \& Navarro 
1999; Abadi et al. 2003; Kawata et al. 2004; Bailin et~al. 
2005; Okamoto et al. 2005; Governato et al. 2006). 
As a benchmark for the semi-cosmological framework, we have 
also analysed the disc galaxy simulations from Bailin et al. 
(2005), which are fully cosmological within a 
$\Lambda$-dominated CDM cosmology (see Bailin et al. 2005 
for details).

In the following, and for the ensemble of the simulated 
galaxies, the stellar mass is derived from the stars
situated within 15~kpc of the centre of the 
stellar mass distribution, while the metallicity and the 
luminosity-weighted age of integrated stellar populations 
and the interstellar medium abundances are measured within 
10~kpc of the centre. Stellar particles in the 
simulated galaxies represent simple stellar populations 
with a given age and metallicity. The photometric properties 
of the simulated galaxy stellar populations were estimated 
using the population synthesis models of Mouhcine 
\& Lan\c{c}on (2003), taking into account both the age and 
the metallicity of each stellar particle. Broad band optical 
properties have been converted into the SDSS photometric 
system as in Fukugita et al. (1996).

%%%%%%%%% STAR FORMATION HISTORIES
\begin{figure}
\includegraphics[clip=,width=0.45\textwidth]{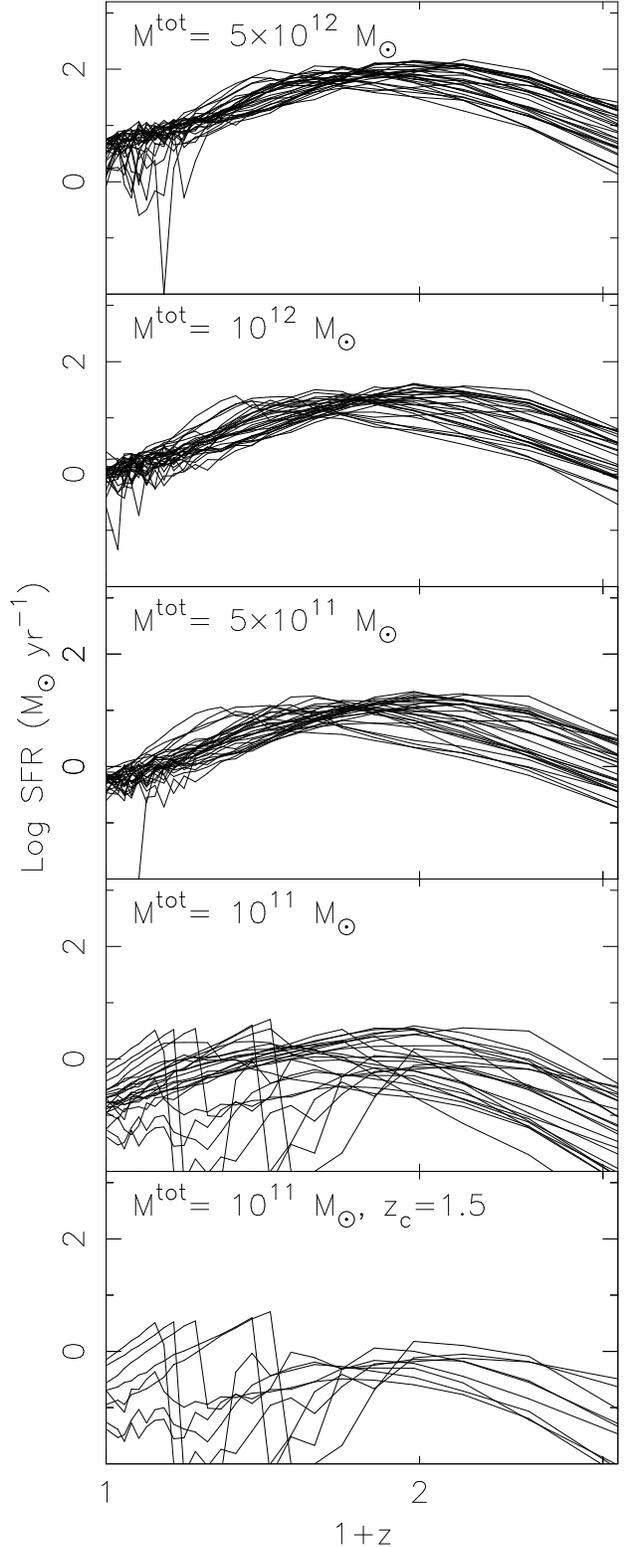}
\caption[]{Star formation rates as a function of 
redshift for the ensemble of galaxies analysed here.}
\label{sfr_hist}
\end{figure}

%%%%%%%%% TOTAL AND STELLAR MASS ASSEMBLY %%%%%%%
\begin{figure*}
\includegraphics[clip=,width=0.45\textwidth]{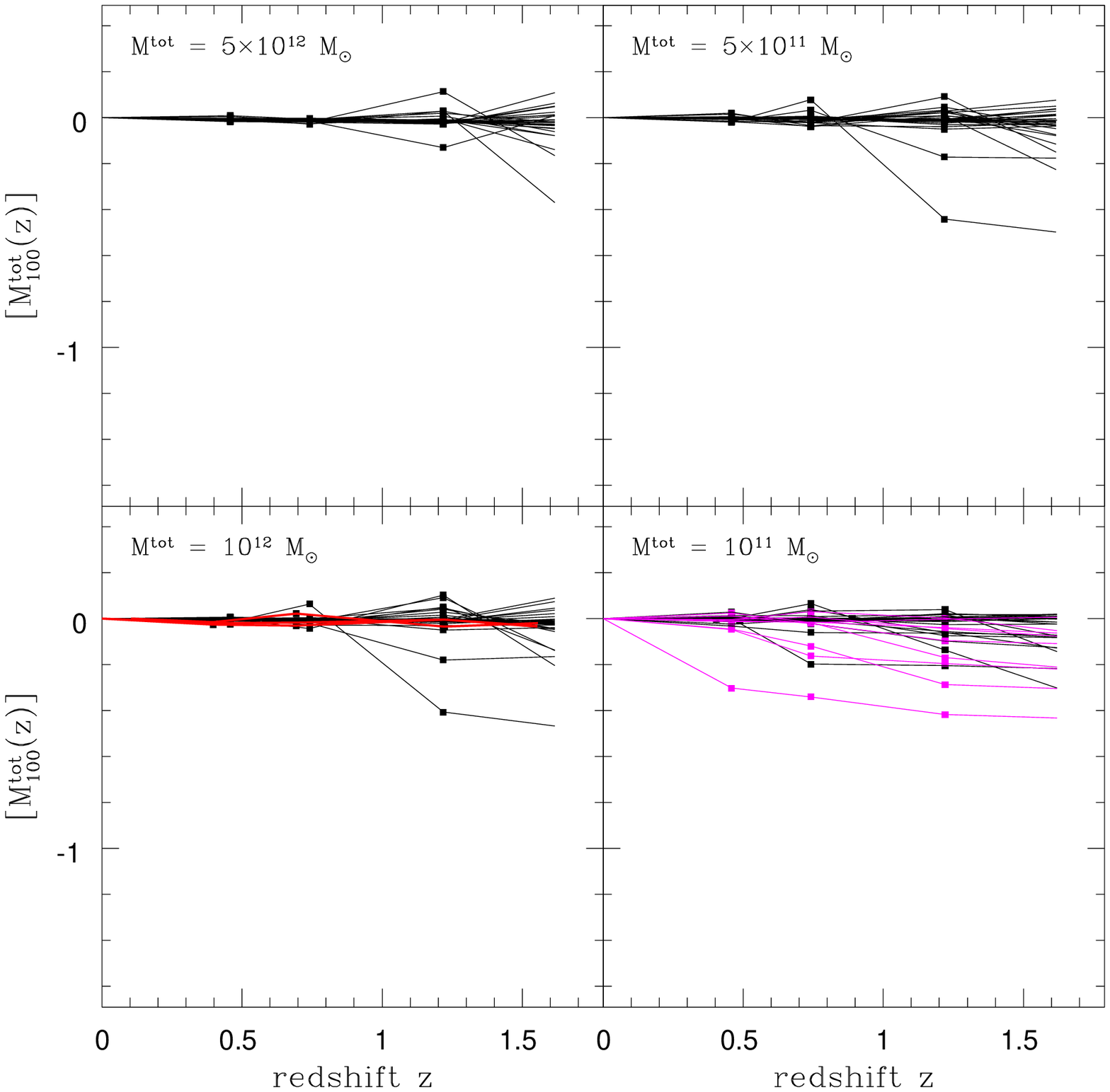}
\includegraphics[clip=,width=0.45\textwidth]{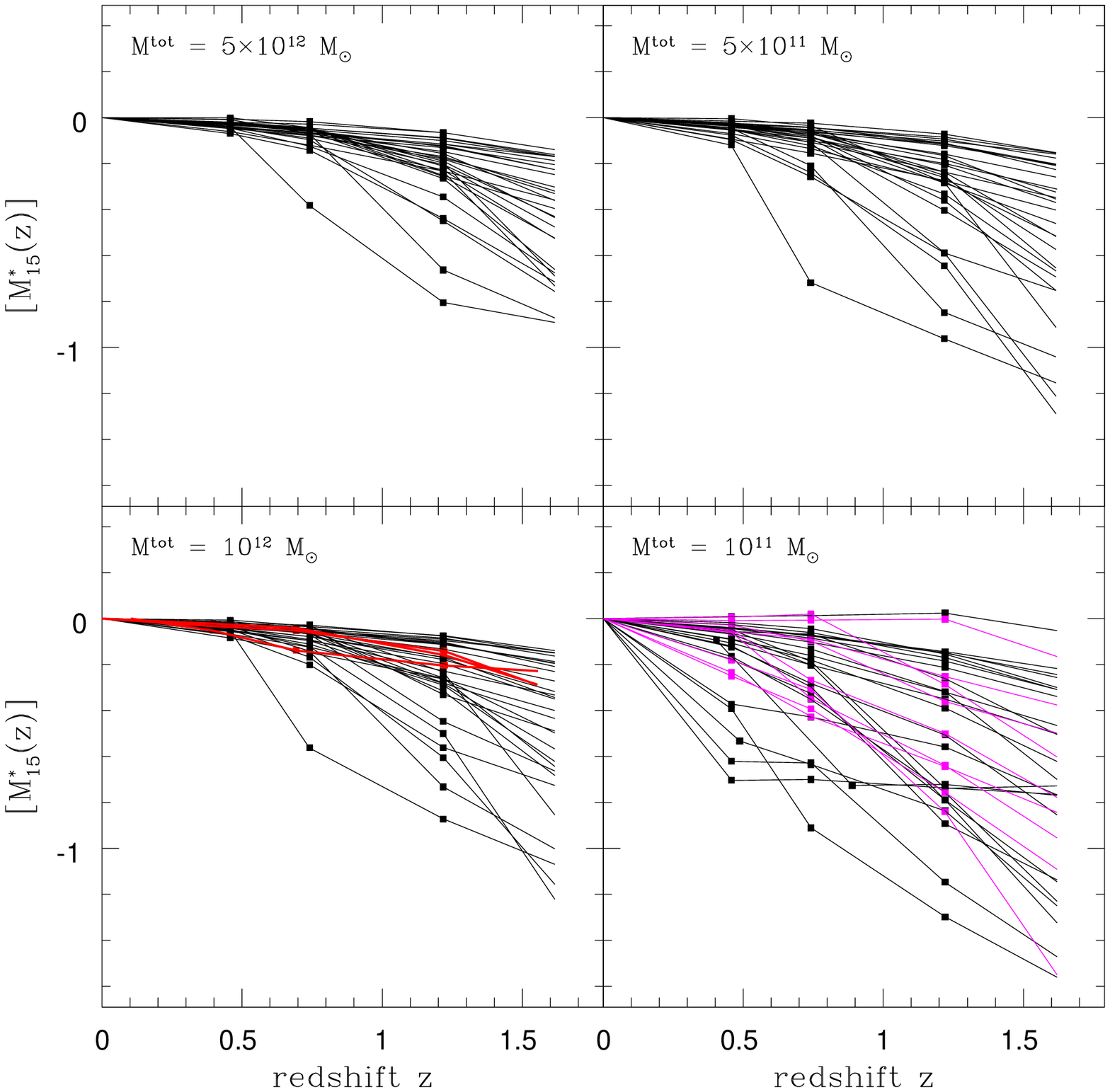}
\caption[]{Left: the redshift evolution of the total mass 
within the 100~kpc central region normalised to the $z=0$ 
value for each subset. The semi-cosmological simulations 
with collapse redshift $z_{\rm c}=2$ are shown as black 
solid lines, those with collapse redshift $z_{\rm c}=1.5$ 
are shown in magenta. The cosmological simulations are shown 
in red. Right: similar to the left panel but for the stellar 
mass within the 15~kpc central region normalised to the 
$z=0$ value.}
\label{mass_assembly}
\end{figure*}

%%%%%%%%%%%%%%%%%%%%

\section{Results}
\label{resl}

\subsection{Galaxy mass assembly}

Figure \ref{sfr_hist} shows the evolution of star formation 
rates for all stars within $r=15$ kpc at $z=0$ as a 
function of redshift for the ensemble of galaxies in our 
synthetic sample. The figure shows that the star formation 
histories of our simulated galaxies are quite diverse. 
%
%As stars form, thermal energy of supernovae is injected 
%into the interstellar medium. The energy is distributed 
%to the surrounding gas particles in our model. The hot gas 
%region expands and the gas density becomes so low that star 
%formation is terminated. 
For massive galaxies, the star formation takes place 
continuously and smoothly. 
This is because the enhancement of the star formation due 
to the deeper potential and the existence of heavy elements 
(leading to rapid radiative cooling) is larger than the 
star formation supression by thermal feedback (see Kawata 
\& Gibson 2003 and Kobayashi et al. 2007 for similar 
analysis). For the majority of these galaxies, the star 
formation rates show peaks around $z\sim 1$ and then decline 
during later cosmic epochs. For low mass galaxies, the star 
formation histories show however striking differences. 
Due to their low metal content and shallower potential, the 
energy feedback from supernovae explosions suppresses more 
efficiently the star formation in low mass systems, leading 
to more bursting star formation histories.

Figure \ref{mass_assembly} shows the redshift evolution 
of the stellar mass (right panel), and the total mass 
(left panel), the latter including both dark and baryonic 
matter within the inner 100~kpc, normalized to their 
value at $z=0$ for the ensemble of simulated galaxies 
ranked by their total mass. Solid black lines show the 
semi-cosmological runs, solid red lines the cosmological 
ones, and solid magenta lines the semi-cosmological 
simulations with collapse redshift $z_{\rm c}=1.5$. 
We note in passing that the assembly histories of both 
the total and the stellar mass in the semi-cosmological 
{\it and} the cosmological simulations are similar. 
The observed consistency is reassuring and leads us to 
conclude that the former are not affected significantly 
by not being fully cosmological.

The figure shows that the assembly of the total mass 
within the inner 100~kpc regions of galaxies is completed 
by $z\ga 1.5$, independently of their total mass. 
On the other hand, the stellar mass in the central 15~kpc 
regions shows more extended assembly histories, with a 
large variety of assembly histories.
%; i.e., at z~$\sim1$,
%approximately half of the baryonic mass in the inner 
%region is gaseous, with a large variety of assembly 
%patterns.
While a fraction of the simulated galaxies have 
assembled their stellar content at early times, typically 
by $z\sim1.5$,
% and have exhausted a significant fraction
%of their gas supply at the present epoch, 
others continue to form stars to much later epochs.
% whereas the galaxies 
%with the collapse redshift z$_{\rm c}~=~1.5$ have their 
%baryonic content dominated by the gaseous component at 
%all redshifts. 
Galaxies with comparable stellar mass can 
have significantly different assembly histories of their 
stellar content, and diverse gaseous contents. As star 
formation does not proceed uniformly in all systems, 
galaxy stellar mass should not necessarily be considered 
a robust indicator of the merging history nor a good 
tracer of the total baryonic galaxy mass.

%%%%%%%%%% STELLAR MASS VS. TOTAL MASS %%%%%%%%%%

\begin{figure}
\centering
\includegraphics[clip=,angle=0,width=0.5\textwidth]
{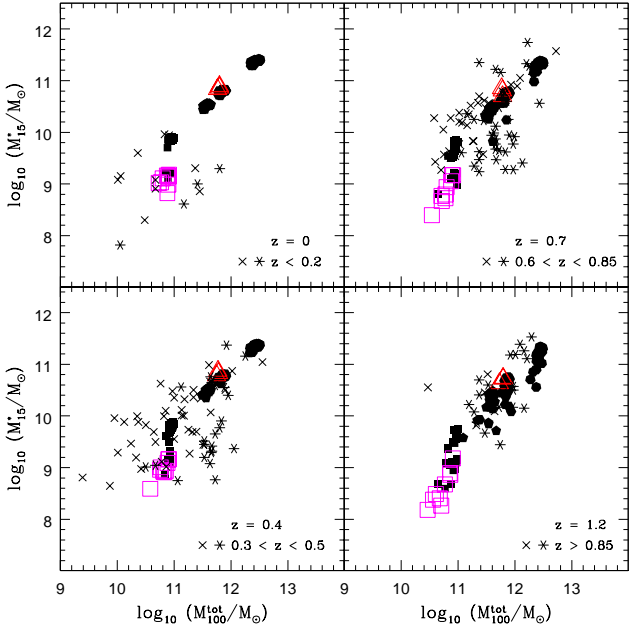}
\caption{The stellar mass against the total mass at 
different redshifts in our simulated sample. 
The semi-cosmological simulations with collapse redshift
$z_{\rm c}=2$ are shown as filled symbols, those with
collapse redshift $z_{\rm c}=1.5$ are shown as larger
empty boxes. The cosmological simulations are shown as
open triangles.
The relationships for simulated galaxies are compared 
with the observed ones in B\"ohm \& Ziegler (2006) and 
Conselice et al. (2005), shown as four and six vertices 
stars, respectively.} 
\label{mtot_ms_obs}
\end{figure}

%%%%%%%

Fig.~\ref{mtot_ms_obs} shows the relationship between
the total mass and the stellar mass of the galaxies in 
our simulated sample at different redshifts. 
The semi-cosmological simulations with collapse redshift
$z_{\rm c}=2$ are shown as filled symbols, those with
the collapse redshift $z_{\rm c}=1.5$ are shown as open 
boxes. The cosmological simulations are shown as open 
triangles.
The predicted relationship is compared to the observed 
data in B\"ohm \& Ziegler (2006) and Conselice et al. 
(2005), shown as four and six vertex stars, respectively. 
A correlation between the stellar and the total mass is 
present from $z\sim 1$ to the present day. The galaxy 
stellar mass increases as a function of the mass of the 
host dark halo, in agreement with the observed correlation. 
The scatter in the model around the predicted relation 
increases with redshift. Galaxies with a total mass of 
${\rm \sim10^{11}~M_{\odot}}$ show a factor of ten 
variation in the stellar mass at all redshifts. 
At a given total mass, simulated galaxies with low stellar 
mass tend to be less evolved than those with more massive 
stellar contents. It is worth to mention that a tighter 
correlation is predicted between the total mass and the 
baryonic mass, where the latter span a more restricted 
range than the stellar mass. 
The predicted correlation is due to the fact that smaller 
systems tend to have a larger fraction of baryons in the 
form of gas at all redshifts (see below), in agreement 
with the results of De Rossi et al. (2007).

%%%%%%%%% MASS-OXYGEN ABUNDANCE / GASEOUS COMPONENT

\begin{figure}
\includegraphics[clip=,angle=90,width=0.5\textwidth]
{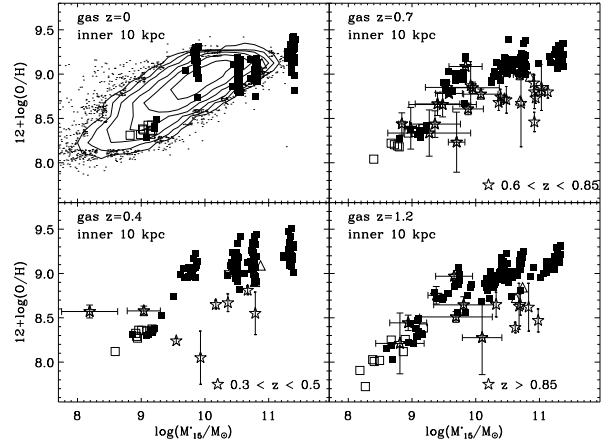}
\caption{The oxygen abundance of the gaseous component 
against the stellar mass, at different redshifts for 
simulated galaxies. The symbols are the same as in 
Fig.~\ref{mtot_ms_obs}. The gas oxygen abundance 
predicted at the present epoch is compared with the 
relation obtained for a large sample of local SDSS 
late-type star-forming galaxies shown as small points
and solid contours. The contours are spaced in number 
density with a factor of 2 between two consecutive 
contours. The results at intermediate redshifts are 
compared with the mass-oxygen abundance relationship 
in Savaglio et al. (2005) and Liang et al. (2006) 
shown as open stars.}
\label{oh_ms_gas} 
\end{figure}

%%%%%%%%%%%%%%%%%%%%%%%%%%%%%%%%%%%%%%%%%%%%%

%%%%%%% GAS BARYONIC FRACTION  VS. STELLAR MASS %%%%%

\begin{figure}
\includegraphics[width=0.5\textwidth]{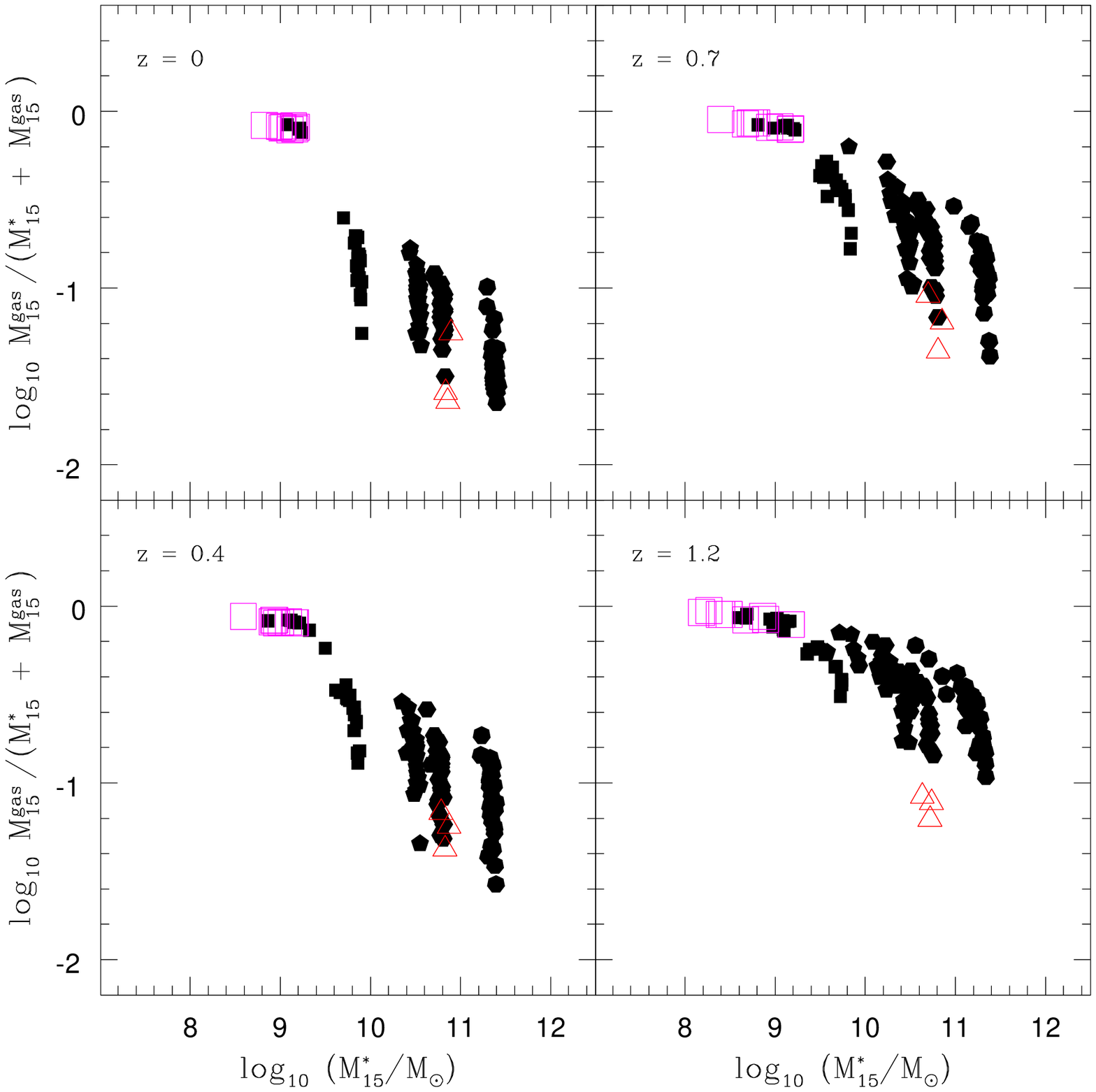}
\caption{The relationship between the gaseous fraction 
and the galaxy stellar mass at different redshifts.
The symbols are the same as in Fig.~\ref{mtot_ms_obs}. }
\label{gf_ms} 
\end{figure}

%%%%%%%

%%%%%%%%%%%%%%%%%%%%%%%%%%%%%%%%%%

\subsection{Gas-phase oxygen abundance vs. stellar mass}

Our simulations follow the enrichment of individual 
chemical elements, so we estimate directly the gas-phase 
oxygen abundance as \doh. 
%We have calculated the oxygen 
%abundance of the gaseous component of a simulated galaxy 
%using the number ratio of oxygen to hydrogen in the 
%gas-phase. 
In Fig.~\ref{oh_ms_gas} we show the relationship between 
stellar mass and oxygen abundance of gas particles for 
galaxies in our simulated sample at different cosmic 
epochs. 
%
% The symbols are the same as in Fig.~\ref{mtot_ms_obs}. 
%
Note that the results for the cosmological simulations 
broadly agree with the semi-cosmological ones in spite 
of the differences between the two frameworks (see 
Sect.~\ref{simul}).

The results at redshift $z=0$ are compared to the 
observed local relation. The local sample used here is 
the sample of star-forming galaxies in the so-called 
``main galaxy sample'' of the SDSS Data Release Four 
(Adelman-Mccarthy et al. 2006). Gas-phase oxygen abundances 
are determined as described in Tremonti et al. (2004).
%For intermediate redshifts, the predicted mass-metallicity 
%relations are compared to the observed ones between 
%$0.4~\leq~z~\leq~1$ from Savaglio et al. (2005) and 
%Liang et al. (2006) shown as open stars, respectively. 
At the present epoch, 
the predicted relationship between the stellar mass and 
the gas-phase oxygen abundance changes as a function of 
the galaxy stellar mass. For the galaxies with stellar 
masses larger than $\sim10^{10}$~M$_{\odot}$, the 
stellar mass-gas oxygen abundance relation is flat, with 
significant scatter, with solar abundances (\doh=8.69; 
Allende Prieto et al. 2001) or above, while galaxies 
with stellar masses lower than $\sim10^{10}$~M$_{\odot}$ 
show lower oxygen abundances, with a sharp transition 
between these two classes of galaxies. Massive galaxies 
with similar stellar masses show a range of gas-phase 
abundances, suggesting that stellar mass is not the only 
driver of the metallicity. The sharply separated 
population of metal-rich galaxies with stellar masses of 
$\sim10^{10}$~M$_{\odot}$ and metal poor galaxies with 
stellar masses of $\sim10^{9}$~M$_{\odot}$ come from 
models with total masses of $M_{\rm tot}=10^{11}$~M$_{\sun}$. 
The separation of the two population is due to the 
difference of the assembly history of galaxies, and the 
strong suppression of star formation in low mass systems
to $z=0$. 

The results at $z=0$ are in broad agreement with 
the local relation for star-forming galaxies. We would like 
to stress that we have not adjusted the absolute value of 
gas-phase oxygen abundance to the observed value, nor have 
we fine-tuned the star formation efficiency to match the 
data. Therefore, it is encouraging that the absolute value 
of the oxygen abundances of the simulated galaxies at the 
present day is consistent with the observed ones. 
The predicted scatter of the gas-phase metallicity at 
a given galaxy stellar mass at $z=0$ is similar to that 
observed for local galaxies in the SDSS. 

As explained in Sect.~\ref{simul}, both the collapse 
redshift $z_{\rm c}=2$ and the spin parameter
$\lambda=0.06$ have been fixed in order to sample only 
the pattern of initial density perturbations, and thus 
the formation history, in the semi-cosmological framework. 
We have released this constraint on the subset at 
M$_{\rm tot} = 10^{11}$~M$_{\odot}$ and completed nine 
runs with collapse-redshift $z_{\rm c} = 1.5$ and 
spin-parameter $\lambda = 0.054$. 
Fig.~\ref{oh_ms_gas} shows that our sub-sample
of M$_{\rm tot} = 10^{11}$~M$_{\odot}$ systems with
$z_{\rm c} = 1.5$ are broadly as evolved as the most 
metal-poor simulations at the same M$_{\rm tot}$ with 
collapse-redshift $z_{\rm c} = 2$. This suggests that 
the feedback from supernovae more easily suppresses 
star formation in galaxies with lower collapse redshift, 
i.e. the galaxies formed from lower $\sigma$ density 
perturbations.

Fig.~\ref{oh_ms_gas} shows that the redshift evolution 
of the relationship between the stellar mass and the gas 
metallicity depends on the mass. For galaxies with stellar 
masses larger than $\sim 10^{10}$~M$_{\odot}$, the level 
of chemical enrichment shows a modest change as a 
function of the redshift, as they have consumed large 
fractions of their gas reservoirs at high redshift. 
Their mass-metallicity relationship remains nearly as 
flat as at the present epoch. As the universe ages, there 
is however a systematic enrichment of the metal content 
of the low stellar mass galaxies, together with the 
increase in their stellar contents by a large factor, 
migrating from stellar masses of $\sim 10^{9}$~M$_{\odot}$ 
to $\sim 10^{10}$~M$_{\odot}$. Due to their extended star
formation histories, masses and metallicities of low mass 
galaxies evolve more slowly than for their high mass 
counterparts. At all redshifts, the slope of the stellar
mass-metallicity relation for low stellar mass galaxies 
tends to be steeper than the relation outlined by massive 
ones. The build up of the transition from the metal-rich 
and high stellar mass galaxies to the metal-poor and low 
stellar mass ones is clear at $z=0.4$. The transition 
from high stellar mass, metal-rich galaxies to low stellar 
mass, metal-poor galaxies gets smoother at higher redshift. 

The predicted stellar mass-metallicity relations at 
intermediate redshifts are in broad agreement with the 
observed ones between $0.4\leq z \leq 1$ from Savaglio 
et al. (2005) and Liang et al. (2006). 
The observational data show hints of a flat stellar 
mass-metallicity relation for massive galaxies up to 
$z\sim 1$, and an increase of the gas-phase metallicity 
as a function of the stellar mass at lower stellar masses. 
The predicted gas-phase oxygen abundances for low stellar 
mass simulated galaxies agree nicely with the observed 
ones, however gas-phase oxygen abundances of simulated 
massive galaxies tend to be systematically higher than 
observed.
The metallicity evolution found in our simulations for 
massive galaxies since $z\sim 1$ is slower than observed.
A better agreement with observations seems to require a 
feedback mechanism to delay the star formation to later 
epochs in massive galaxies in order to prevent those 
galaxies from enriching their gas to a high metallicity 
at high redshifts.
Note however that (at least) part of the offset between 
observed and simulated gas-phase abundances for massive 
galaxies could be due to systematic errors affecting 
the empirical measurements. The procedure used to derive 
gas-phase metallicities for intermediate redshift galaxies, 
i.e., the strong line method in the absence of 
temperature-sensitive emission lines, has been suspected 
to involve systematic errors, especially at the 
high-metallicity end (e.g., Kennicutt et al. 2003; 
Bresolin et al. 2004). In addition, oxygen abundances 
estimated using different calibrations of the strong 
line method may differ by factors of up to $\sim4$ 
(Ellison \& Kewley 2005). 

If the star formation efficiencies are mass-dependent,
more massive galaxies convert larger fractions of their 
gas into stars than low mass counterparts, and then 
have shorter timescale for their chemical evolution; 
therefore this could lead to the observed trend between 
stellar mass and gas-phase oxygen abundance.
Fig.~\ref{gf_ms} shows the relationship between 
the baryonic gas fraction -- i.e., the ratio between the 
gaseous mass and the baryonic mass -- and the stellar 
mass at different redshifts. 
%
% The symbols are the same as in Fig.~\ref{mtot_ms_obs}.
%
In our simulations, the baryonic gas fraction correlates 
with the stellar mass, in agreement with the observations 
of the local disc galaxies (e.g. Rosenberg et al. 2005). 
Low mass galaxies have larger gas fraction, implying 
that they have converted lower fractions of their 
interstellar gas into stars than galaxies with higher 
mass, and they are less chemically enriched.
%, and require consequently a much longer timescale 
%to enrich their interstellar gas to the metallicities 
%of galaxies with higher stellar mass.
%
Even though a trend is present, which gets shallower at 
higher redshifts, galaxies of a given stellar mass span 
a wide range of gaseous-to-stellar mass ratio as a 
consequence of the large variety of their assembly 
histories. Note that the cosmological simulations tend 
to have low gas fraction. This can be traced to not 
having adopted the strong supernova feedback model 
of Brook et al. (2004), but adopting pure thermal 
feedback in these simulations. The weak feedback leads 
to more efficient consumption of the gas component, 
and their gas fraction remains low.

Our results agree with recent investigations that have 
attempted to place the stellar mass versus metallicity 
relation into an hierarchical context, using numerical 
simulations. De Rossi et al. (2007) used cosmological 
hydrodynamical simulations to reproduce roughly the 
local relation as a consequence of the variation of 
star formation efficiency with stellar mass. Similarly, 
Brooks et al. (2007) argue that galaxy mass loss does 
not directly suppress the metallicities of low mass 
galaxies, but that supernova feedback leads to 
low star formation efficiencies in those galaxies, which 
leads to low chemical abundances. Using high resolution 
cosmological simulations of high redshift galaxies, 
Tassis et al. (2006) have found that gas outflows 
escaping the host galaxy halo are not required to 
explain the observed scaling relations of dwarf galaxies. 
Again, scaling relations in simulated galaxies similar 
to those observed arise due to increasingly inefficient 
conversion of gas into stars in low mass galaxies rather 
than outflows.

Finlator \& Dav\'e (2007) have compared the observed 
stellar mass versus metallicity relation of star 
forming galaxies at $z\approx 2$ from Erb et al. (2006) 
with predictions from cosmological hydrodynamical 
simulations to argue however that outflows are required 
not only to suppress star formation but also to lower 
galaxy metal content. They claim that a feedback model 
where the outflow velocity scales as the escape velocity 
is necessary to reproduce the slope and normalization 
of the observed relation at $z\approx 2$. 
In this scenario, the scatter in the relation is due 
primarily to the dilution timescale, representing the 
timescale for a galaxy to return to an equilibrium
metallicity (reflecting the enrichment balance between 
star formation and gas accretion) after a 
metallicity-perturbing interaction, compared to the 
dynamical timescale. 
%
%In our simulations, the origin 
%of the scatter lies in the differing merging histories 
%among the galaxies with similar masses, and is found 
%to be similar to that observed. 
%
However a potential concern here is that (i) the 
metallicity indicator used by Erb et al. (2006) may 
saturate for high metallicity galaxies, and (ii) the 
calibration used by those authors may underestimate 
oxygen abundances (see Liang et al. 2006 and 
Kennicutt et al. 2003 for more details). 
The normalization of the observed relation at $z\approx 2$ 
is still uncertain (see Erb et al. 2006 for more details). 
Interestingly, the slope of the stellar mass vs. 
metallicity relation predicted in their no-wind case is 
in fair agreement with observations, although simulated 
galaxies have higher abundances at a given stellar mass 
than estimated observationally. 
Finlator \& Dav\'e (2007) have not followed their 
simulations to the present epoch, where the observed 
stellar mass - metallicity relation is much less 
affected by systematic effects.

\subsection{Stellar metallicity/age vs. stellar mass}

%%%%%%%%% MASS-METALLICITY / STELLAR COMPONENT

\begin{figure}
\includegraphics[clip=,angle=90,width=0.5\textwidth]
{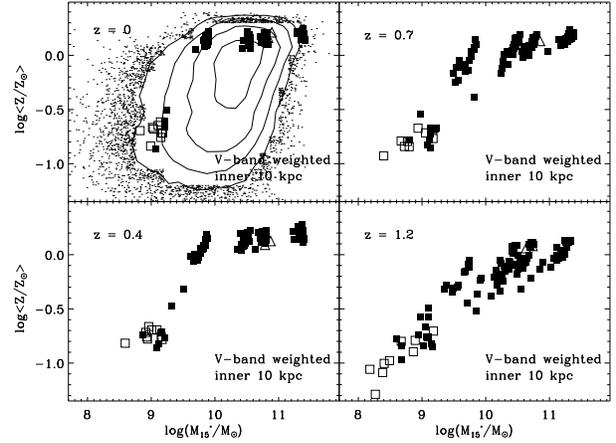}
\caption{The V-band luminosity-weighted mean total 
metallicity of the integrated stellar populations against 
the integrated stellar mass, at different cosmic epochs. 
The symbols are the same as in Fig.~\ref{mtot_ms_obs}. 
The results at $z=0$ are compared with the relation 
between the stellar mass and the total metallicity of 
local late-type galaxies in the SDSS shown as a solid 
contours and small points. The contours are spaced in 
number density with a factor of 2 between two consecutive 
contours. See the text for more details.}
\label{zs_mstar}
\end{figure}

%%%%%%%%% M/L VS. COLOUR %%%%%%%%%%%%
\begin{figure*}
\centering
\includegraphics[clip=,width=0.45\textwidth]{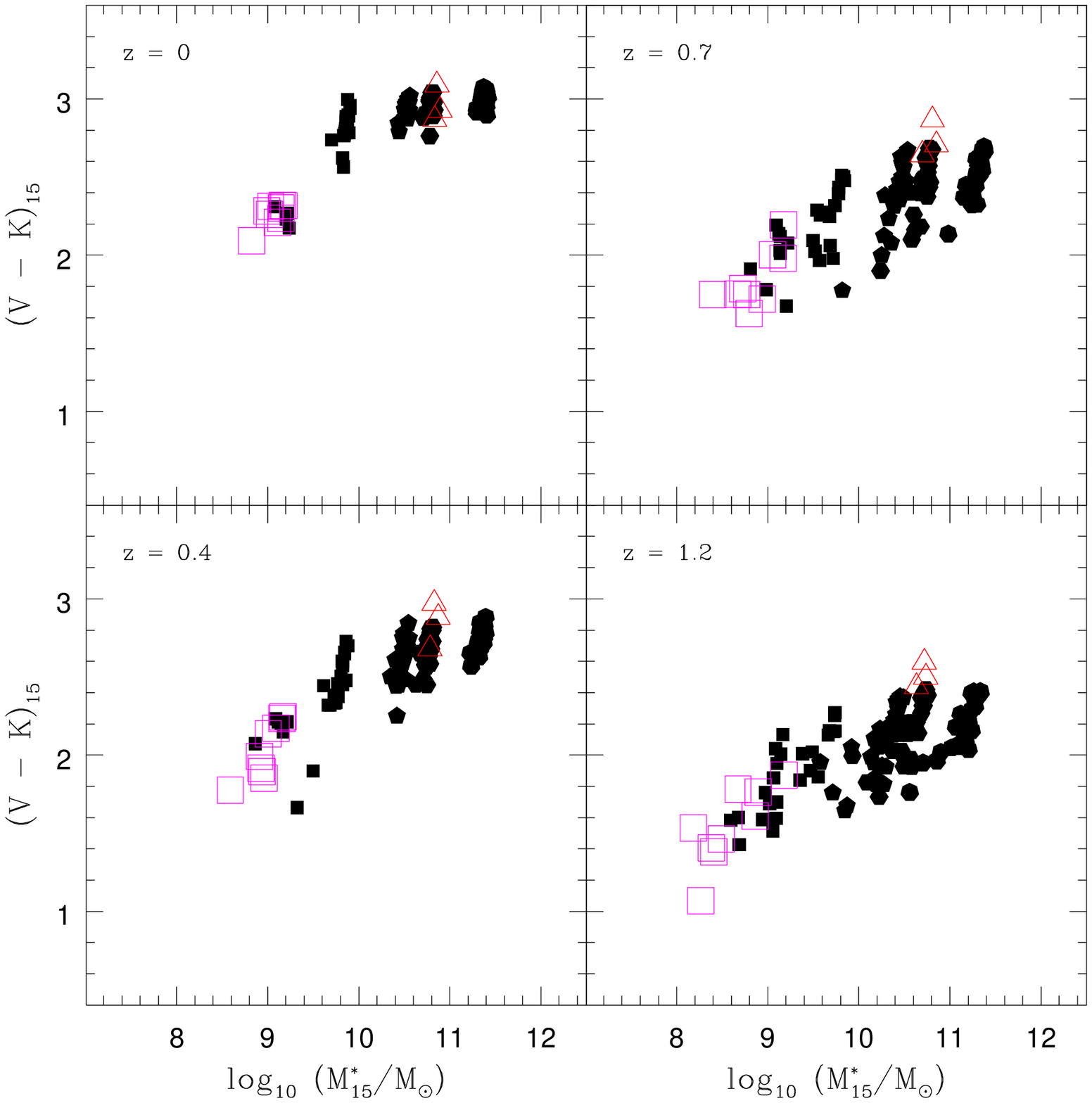}
\includegraphics[clip=,width=0.45\textwidth]{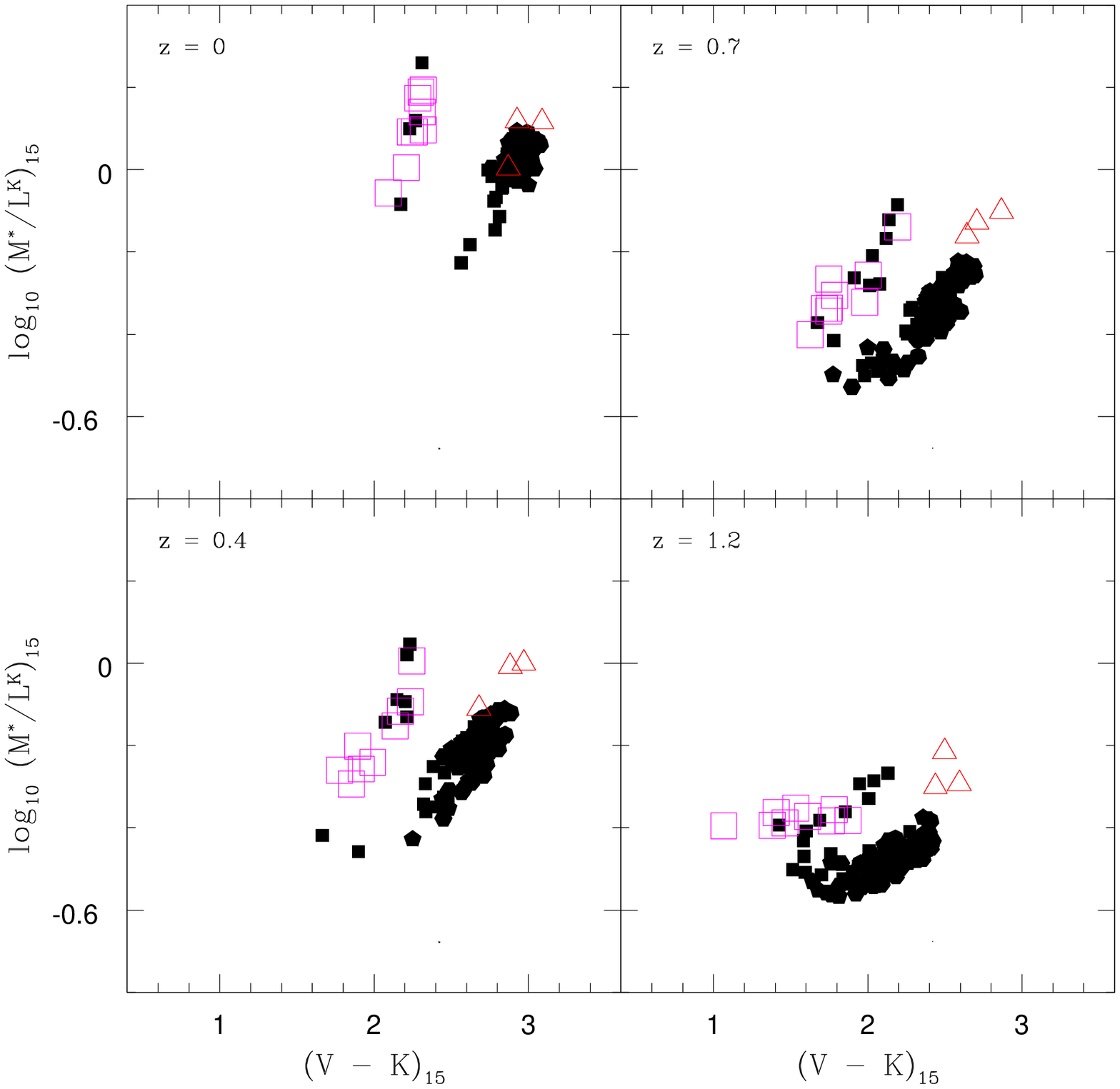}
\caption{The relationships between the integrated stellar 
population (V-K) colour and the stellar mass (left panel) 
and the K-band stellar mass-to-light ratio (right panel) 
in our simulated sample at different redshifts. 
The symbols are the same as in Fig.~\ref{mtot_ms_obs}.}
\label{MLratio_col}
\end{figure*}

%%%%%%%%% MASS-AGE

\begin{figure}
\includegraphics[clip=,angle=90,width=0.5\textwidth]
{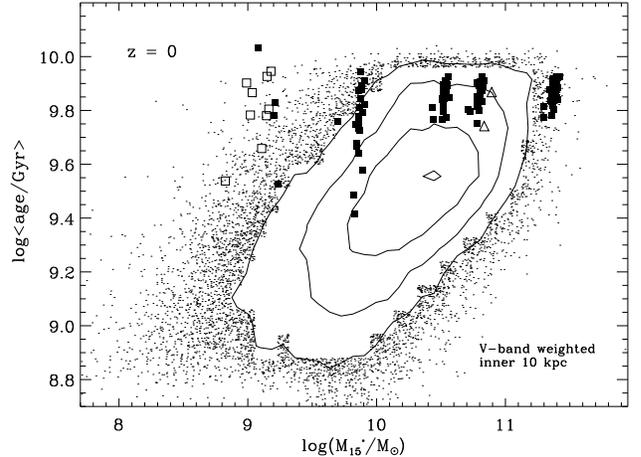}
\caption{The V-band luminosity-weighted mean age for the
integrated stellar populations, against the stellar mass 
at redshift $z=0$. The symbols are the same as in 
Fig.~\ref{mtot_ms_obs}. The results are compared with the 
relation between the stellar mass and the age of local 
late-type galaxies in the SDSS. The contours 
are spaced in number density with a factor of 2 between two 
consecutive contours.}
\label{agestar_mstar}
\end{figure}

%%%%%%%%% STELLAR-TO-TOTAL MASS RATIO VS. AGE

\begin{figure}
\includegraphics[width=0.5\textwidth]{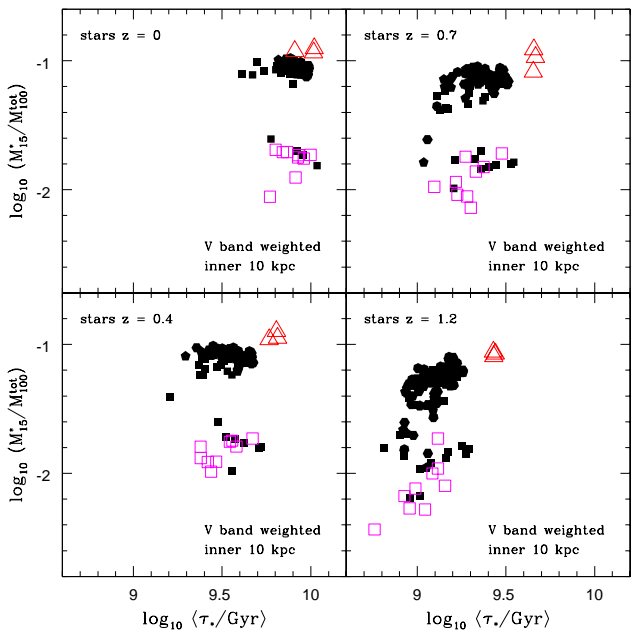}
\caption{The stellar-to-total mass ratio against the 
V-band luminosity-weighted mean age for the integrated 
stellar populations. The symbols are the same as in 
Fig.~\ref{mtot_ms_obs}. }
\label{massratio_age}
\end{figure}

Ages and metallicities of stellar populations are powerful 
tracers of galaxy star formation and chemical enrichment 
histories, and offer a way of constraining their assembly 
histories. Fig.~\ref{zs_mstar} shows the relationship 
between the V-band luminosity-weighted total metallicity 
and the integrated stellar mass at different redshifts. 
%
% The symbols are the same as in Fig.~\ref{mtot_ms_obs}.
%
The results at redshift $z=0$ are compared to the 
observed local relation. The local sample used here was
selected from the ``main galaxy sample'' of the SDSS Data 
Release Four. The models presented here are designed for 
late-type galaxies. For a meaningful comparison, we have 
split this sample into nominally early and late-type 
galaxies. The (inverse) concentration index, defined as 
the ratio of the radii enclosing 50\% ($R_{50}$) and 90\% 
($R_{90}$) of the Petrosian $r$-band galaxy light, is used 
as a proxy for galaxy morphology indicator
(Shimasaku et al. 2001; Strateva et al. 2001). 
Baldry et al. (2006) have found that a concentration index 
around $\sim 0.4$ is a natural dividing line between red 
and blue galaxy populations for all stellar masses. 
We selected late-type galaxies from the ``main galaxy sample''
by selecting galaxies with concentration index greater than 
0.4. Stellar metallicities are determined as described in 
Gallazzi et al. (2005).
The predicted relationship between the luminosity-weighted 
stellar metallicity and the integrated stellar mass at the 
present epoch is in broad agreement with the observed 
local relation for late-type galaxies over $\sim$3 orders 
of magnitude in stellar mass. 

The general behaviour of the predicted relationship between
stellar metallicity and stellar mass
at the different redshifts is similar to the evolution 
of the relation for the gas-phase oxygen abundance,
with the predicted stellar metallicities less dispersed 
than the gas-phase oxygen abundances at a 
given galaxy stellar mass. The predicted local relationship 
shows a transition from massive galaxies, with a flat 
metallicity-mass relation, to low stellar mass galaxies, 
with a steeper relation. The redshift evolution of the 
stellar metallicity-mass relation is mainly driven by 
the low mass systems, in agreement with the results of 
De Rossi et al. (2007) who found that systems with stellar 
masses smaller than a few times 
${\rm \sim\,10^{10}\,M_{\odot}}$ are responsible for the 
evolution of this relation at least from $z\approx 3$.
Massive galaxies in the simulations assembled their 
stars earlier (see Fig.\,\ref{mass_assembly}), making 
the relation steeper for low mass systems than for their 
massive counterparts. 
%Galaxies with stellar masses 
%${\rm \sim\,10^{10.5}\,M_{\odot}}$ span a large range of 
%oxygen abundances. 
%

The left panel of Fig.~\ref{MLratio_col} shows the 
relationship between the integrated stellar population 
(V-K) colour and the stellar mass. 
% 
% The symbols are the same as in Fig.~\ref{mtot_ms_obs}. 
%
The colours are estimated using the luminosities of the 
integrated stellar populations within the central regions 
of simulated galaxies. Red galaxies in the simulated 
sample are predominantly massive, in agreement with 
observations (e.g., Borch et al. 2006), as the most 
massive simulated galaxies have the most metal-rich 
stellar populations (see below). The relationship between 
integrated stellar population colours and stellar mass 
is present up to $z\sim 1$, although the scatter in the 
correlation is more substantial at earlier cosmic epochs. 
For present-day galaxies in our simulated sample, the 
(V-K) colours seem to 
flatten at stellar masses ${\rm \sim10^{10}>M_{\odot}}$. 
This flattening gets weaker at higher redshifts. 

The fraction of the baryonic mass converted from gas 
into stars in the simulated galaxies (not shown here)
is found to 
correlate with the stellar population colours at all 
redshifts -- i.e. gas rich galaxies are on average the 
bluer ones, in agreement with the observed local 
correlation between the atomic gaseous-to-stellar mass 
ratio and the optical/near-infrared colour (Kannappan 
et al. 2004; see also Geha et al. 2006). 
The relationship is driven by the correlation between 
the baryonic gas fraction and the stellar metallicity, 
in agreement with that inferred from the observed 
correlation between the metallicity-sensitive 
near-infrared colour and the gaseous mass fraction 
(Galaz et al. 2002). This also implies that the lower level 
of chemical evolution of the stellar component of low 
mass galaxies is primarily due to their low efficiency 
at converting gas into stars, as discussed above.

The right panel of Fig.~\ref{MLratio_col} shows the
relationship between the K-band stellar mass-to-light
ratio and the integrated stellar population (V-K) colour. 
%
% The symbols are the same as in Fig.~\ref{mtot_ms_obs}. 
%
Although with a large scatter,
% a trend where redder 
% have higher stellar mass-to-light ratio is 
%present. 
galaxies with low stellar mass and high baryonic gas 
fraction, tend to have lower mass-to-light ratio and 
bluer colours than those with high stellar mass and low 
gaseous fraction, in agreement with recent observational 
results (Galaz et al. 2002). We also found that the mass 
of the gaseous component in the simulated galaxies does 
not correlate with the stellar mass-to-light ratio in 
agreement with observations of local disc galaxies 
(Rosenberg et al. 2005). 
The relationship between integrated stellar population 
colour and stellar mass-to-light ratio is present up 
to $z\sim1$, getting shallower at high redshift. 
The predicted correlation between the stellar 
mass-to-light ratio and colour is consistent with the 
suggestion that the stellar mass-to-light ratio varies 
along the Tully-Fisher relation (McGaugh \& de Blok 1997). 
The figure illustrates that
% the large diversity of the 
%evolutionary states of galaxy integrated stellar 
%populations -- i.e., 
galaxies with a given stellar 
mass-to-light ratio span a large range of integrated 
stellar population colours. Each of the two parallel 
sequences in the mass-to-light ratio versus colour 
diagram are populated by galaxies with comparable stellar 
and gas-phase metallicities, but their luminosity-weighted 
ages range from $\sim3$ to $\sim10$~Gyr (see below). 
% The observed scatter of the mass-to-light ratio versus 
% colour relation might thus be, at least partially, driven 
% by variation of stellar metallicity among galaxies. 

Fig.~\ref{agestar_mstar} shows the $z=0$ relationship 
between the stellar mass and the V-band luminosity weighted 
mean age of integrated stellar populations for our sample 
of simulated galaxies. 
%
% The symbols are the same as in Fig.~\ref{mtot_ms_obs}. 
%
The predictions are compared to the relationship derived 
for the local late-type galaxies selected from the SDSS 
Data Release Four. 
We find no relation between the luminosity weighted ages 
and the stellar mass. Large galaxies with stellar masses 
larger than $\sim10^{10}$~M$_{\odot}$ have ages ranging 
from $6$~Gyr to $9$~Gyr, in agreement with the ages of 
SDSS galaxies with similar stellar masses. For low stellar 
mass galaxies in our simulated sample, the mean ages are 
more dispersed than for the massive ones, ranging from 
$\sim3$ to $\sim 10$~Gyr, older than what is estimated 
for local galaxies. 

% For simulated galaxies, since we fixed the collapse 
% redshift, the gas cools and collapses early and 
% independently of the galaxy total mass.
Fig.~\ref{massratio_age} shows the relationship between 
the age and the ratio of the integrated stellar mass to 
the total mass, the latter including both dark and 
baryonic matter within the inner 100~kpc, at different 
redshifts. As shown in Fig.~\ref{mass_assembly} the 
total mass in the inner 100~kpc does not change 
substantially since $z\sim1.5$, and the predicted 
redshift evolution of the stellar-to-total mass ratio 
is driven mainly by the variation of the stellar mass. 
At all redshifts, while the stellar-to-total mass ratio 
covers a wide range, the integrated stellar populations 
tend to be dominated by similarly old stars 
independently of galaxy stellar and/or total mass. 
Erb et al. (2006) have suspected the presence of a 
correlation between the galaxy dynamical-to-stellar 
mass ratio and the stellar population age at $z\sim2$. 
From the Erb et al. (2006) data (see their Fig.~7), it
seems that galaxies with low stellar mass tend to have 
large dynamical-to-stellar mass ratio, indicating that 
those galaxies have recently begun forming stars. 
The so-called downsizing, in which the stellar mass 
of the galaxies undergoing active star formation become 
progressively lower as later epochs (Cowie et al 1996), 
is not seen in our simulations, i.e., there are insufficient
numbers of younger stellar populations in galaxies with 
low stellar masses in our simulations (see also Kobayashi 
et al. 2006 for a similar conclusion). This age problem,
as outlined above, signals two problems in our simulations. 
First, there is an excess of old galaxies with stellar 
masses of $\sim10^{10}$~M$_{\odot}$. 
They correspond to galaxies that show high metallicity 
already at high redshift as seen in Fig.\,\ref{zs_mstar}. 
At the present epoch, they are dominated by old and metal 
rich stellar populations. Their star formation should be 
more suppressed at high redshift, and more self-regulated 
since $z\sim1$. The feedback from supernovae should be 
more effective for these galaxies, preventing gas from 
collapsing and cooling in the early stage of the hierarchy,
in order to ensure that most of the stellar content of 
galaxies does not form early. This has already been 
recognised and invoked in previous numerical studies of 
the angular momentum of galaxies (e.g.: Weil et al. 1998; 
Steinmetz \& Navarro 1999; Sommer-Larsen et al. 2003). 
Second, galaxies with stellar masses of 
$\sim10^{9}$~M$_{\odot}$ are too old. This is due to 
an overly strong suppression of star formation for these 
galaxies. These galaxies form very few stars at the 
present epoch, and are dominated by old stars. 
Note that the luminosity-weighted stellar metallicities 
for these galaxies are consistent with the observed ones 
at $z=0$ (Fig.~\ref{zs_mstar}). 
Their gas-phase metallicity is low (Fig.~\ref{oh_ms_gas}), because 
there is an excess of gas remaining at the present epoch. 
Solving this problem requires a mechanism which maintains 
a moderate amount of star formation since $z\sim1$, and 
blows out a fraction of the gas or suppresses the gas infall. 
A combination of stronger supernovae feedback and the 
UV background radiation (e.g., Efstathiou 1992) may help 
to solve this problem. Alternatively, a mass-dependence 
of the star formation efficiency could also help to 
alleviate the age problem.

%%%%%%%%%%%%%%%%%%%%%%%%%%%%%%%%%%%%%%%%%

\subsection{The Colour-Magnitude Relation}

%%%%%%%%% COLOUR-MAGNITUDE RELATION 

\begin{figure}
\includegraphics[width=0.5\textwidth]{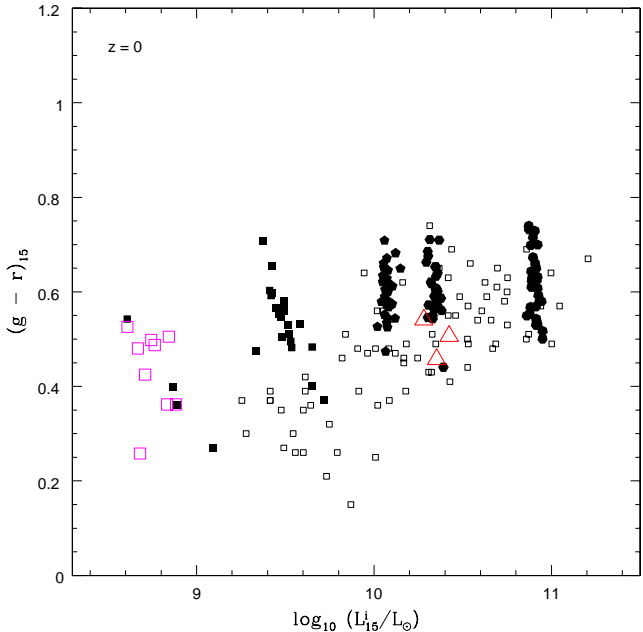}
\caption{The Colour-Magnitude relation for the simulated 
galaxies at $z=0$. The symbols are the same as in 
Fig.~\ref{mtot_ms_obs}. The predicted Colour-Magnitude 
relation is compared with the observed relation for a 
sample of local disc-dominated galaxies selected from the 
SDSS Pizagno et al. (2005), shown as empty small boxes.}
\label{cmr}
\end{figure}

To illustrate a further signature of what is likely to 
be a problem for our simulated sample, Fig.~\ref{cmr} 
shows the relationship between the ($g-r$) colour 
and the $i$-band luminosity for the simulated galaxies 
at the present epoch, compared to the observed relation 
for a sample of disc-dominated galaxies, selected by 
Pizagno et al. (2004) from the SDSS with a disc-to-total
luminosity ratio larger than $0.9$, shown as empty 
squares. The other symbols are the same as in 
Fig.~\ref{mtot_ms_obs}. Since we have not applied any 
treatment for internal extinction in our simulations, 
we have compared the predicted colour-magnitude relation 
to the sample of Pizagno et al. (2004) for which the 
internal dust reddening was corrected for as in 
Tully et al. (1998), rather than to the low redshift 
New York University Value-Added Galaxy Catalogue 
(Blanton et al. 2005) for which magnitudes are not 
extinction-corrected. 

The integrated colour of local late-type galaxies is 
correlated with the absolute luminosity, in the sense 
that bright late-type galaxies tend to be redder than 
the faint ones (e.g. Tully et al. 1982; Bothun et al. 
1984; Peletier \& de Grijs 1998). The colours of simulated 
luminous galaxies are 
in agreement with observed ones, as a consequence of 
the agreement between predicted and observed ages and 
stellar metallicities of massive galaxies. The scatter 
of the integrated colours of simulated galaxies, 
reflecting the differing formation histories which have 
been sampled, agrees with the observed scatter for the 
disc-dominated galaxy colours. Since it is difficult 
to assign morphological types to galaxy models, our 
simulated galaxy sample cannot be trimmed of galaxies 
morphologically earlier than those selected by 
Pizagno et al. (2005). With this caveat, we conclude 
that the colours and the scatter of bright galaxies 
in the simulated colour-magnitude relation are in  
agreement with observations.

Although the predicted colour scatter for the low 
stellar mass simulated galaxies is comparable to what 
is observed, faint galaxies in our simulated sample 
show redder integrated ($g-r$) colours than observed 
for galaxies with similar luminosities, i.e. the 
simulated colour-magnitude relation is shallower than 
observed. 
The difference between the observed and the simulated 
integrated stellar population colours for faint galaxies 
is mainly driven by the differences among the mean ages 
of the stellar populations (Fig.\,\ref{zs_mstar}). 
%
%Although faint galaxies in 
%our simulated sample exhibit stellar and gaseous 
%metallicities in agreement with what is observed for 
%local galaxies with similar stellar mass, the former 
%are older, resulting in redder colours. 
Again, a mechanism to self-regulate the star formation 
in low mass galaxies to later epochs seems necessary to 
alleviate the discrepancy between the measured and the 
predicted stellar population ages and the colour-magnitude 
relation. The agreement found between 
simulated and observed stellar and gaseous metallicities 
at the present epoch, and at earlier cosmic epochs for 
low mass galaxies, suggests that the rate at which stars 
form in simulated galaxies might not be dramatically 
different from what it should be.

\section{Summary}
\label{concl}

We have presented an analysis of the chemical properties 
over the last half of the age of the Universe for a 
sample of 112 simulated late-type galaxies, with stellar 
masses larger than $\sim 10^9$~M$_{\odot}$, formed under 
the hierarchical clustering scenario. 
The simulations include hydrodynamics, star formation, 
metal-dependent cooling, supernova feedback, and a 
chemical enrichment of the interstellar medium that 
relaxes the instantaneous recycling approximation. 
Stellar feedback is taken into account by injecting 
energy into the gas that surrounds regions of recent 
star formation. We compared the observed properties 
of the simulated galaxies with the recent observational 
data at various redshifts. A particular emphasis has 
been placed upon the relationship between the stellar 
mass and the metallicity of both the gaseous and the 
stellar components. The main results can be summarised 
as follows. 

\begin{itemize}
\item[-]
In the hierarchical clustering scenario, even after the 
system has collapsed, the histories of the growth of 
the stellar mass in the central region can be different 
(Fig.~\ref{mass_assembly}) due to the variety in the 
minor merger and gas fueling histories, which also leads 
to the diverse fraction of the gas-to-stellar mass ratio 
among the galaxies with similar stellar or total mass
(Fig.~\ref{gf_ms}).

\item[-]
The stellar mass in the inner region correlates well 
with the total mass of galaxies up to $z\sim 1.2$ 
(Fig.~\ref{mtot_ms_obs}), which broadly agrees with 
the observed relation, especially for higher mass 
(${\rm M_{tot}}>10^{10}$ M$_{\sun}$) galaxies. 
Lower mass galaxies have greater scatter. 
For some simulated galaxies, star formation is heavily 
suppressed and leads to lower stellar mass compared 
with the total mass.

\item[-]
We have analyzed the metallicity for both stellar and gas 
components. Both are correlated with the stellar mass up 
to $z\sim 1.2$, and show an agreement with recent 
observations (Figs.~\ref{oh_ms_gas} and \ref{zs_mstar}). 
This relation arises from the fact that higher mass galaxies 
convert gas more efficiently into stars.

\item[-]
Higher mass galaxies reach a high metallicity earlier.
As a result, the slope of the mass-metallicity relation 
becomes flatter with decreasing redshift
(Figs.~\ref{oh_ms_gas} and \ref{zs_mstar}). 
Lower mass galaxies (${\rm M_{tot}}<10^{10}$ M$_{\sun}$) 
evolve chemically more slowly, and have lower metallicity 
than that predicted from the extrapolation of the 
mass-metallicity relation for higher mass galaxies at lower 
redshifts, which is also consistent with the observational 
data.

\item[-]
The predicted scatter in the mass-metallicity relation 
also reproduces qualitatively the observed scatter 
(Figs.~\ref{oh_ms_gas} and \ref{zs_mstar}). This suggests 
that the observed scatter could be due to the difference 
in mass assembly histories, a natural prediction of the 
hierarchical clustering scenario (Fig.~\ref{mass_assembly}). 

\end{itemize}

The primary disagreement between our simulations and 
observations is as follows:

\begin{itemize}
\item[-]
Our simulations predict almost no correlation between the 
stellar age and the stellar mass, which cannot explain the 
observed significantly younger population in lower mass 
galaxies, e.g. stellar masses lower than 
${\rm \sim 10^{9} M_{\odot}}$ (Fig.~\ref{agestar_mstar}). 
As a result, the simulations also predict too flat of a 
colour-luminosity relation (Fig.~\ref{cmr}).
\end{itemize}

This suggests that the simulations need a mechanism to 
sustain a low level of star formation activity and also 
keep the stellar and gas metallicity low. 
A combination of stronger supernovae feedback and the 
UV background radiation may help to cause such self-regulated 
star formation in the lower mass galaxies. This demonstrates 
that the quantitative comparisons made in the paper are 
invaluable to improve numerical simulation models, which 
eventually will aid in completing our understanding of the 
physical processes governing galaxy formation and evolution. 

\begin{acknowledgements}

We acknowledge the Center for Computational Astrophysics, 
CfCA, of the National Astronomical
Observatory, Japan (VPP5000 was used), the
Institute of Space and Astronautical Science 
of Japan Aerospace Exploration Agency, and
the Australian Partnerships for Advanced
Computing, the Swinburne Supercomputer Facility support team, 
and the Commonwealth Cosmology Initiative.  
AR thanks Michael Blanton for helpful 
discussions, and the Liverpool John Moores University 
Astrophysics Research Institute, the University of Central 
Lancashire Centre for Astrophysics, and the Osservatorio 
Astronomico di Palermo for their kind hospitality.
DK acknowledges the KITP NSF grant, PHY05-51164.

\end{acknowledgements}

%%===============================================================

\end{document}